\documentclass[ %technote,
twoside
conference
]{IEEEtran}
\usepackage{cite}
\usepackage[cmex10]{amsmath}
\usepackage{enumerate}
\usepackage{bbm}
\usepackage{amsmath}
\usepackage{amsfonts}
\usepackage{graphicx}
\usepackage{epstopdf}
\usepackage{caption2}
\usepackage{subfigure}
\usepackage{tabu}

\begin{document}
	\title{A fixed-point model for semi-persistent scheduling of vehicular safety messages  }
	\author{
		\IEEEauthorblockN{Xu Wang\IEEEauthorrefmark{1}, Randall A. Berry\IEEEauthorrefmark{1}, Ivan Vukovic\IEEEauthorrefmark{2}, Jayanthi Rao \IEEEauthorrefmark{2}} 
		
		\IEEEauthorblockA{\IEEEauthorrefmark{1}EECS Department, Northwestern University,} 	 
		\IEEEauthorblockA{\IEEEauthorrefmark{2}Ford Motor Company} 
	}

	\date{}
	\maketitle	
	\begin{abstract}
		In this paper, we focus on the performance analysis of a semi-persistent scheduling scheme for vehicular safety communications, motivated by the Mode 4 medium access control protocol in 3GPP Release 14 for Cellular-V2X.  An analytical model is built and a fixed point method is used to calculate the collision probability and average delay in both fully connected and partially connected cases under the assumption of perfect PHY performance. We use Monte Carlo simulation to verify the results obtained in the analytical model. The simulation results show that our analytical model can give a good estimation of the collision probability and average delay. We verify that a trade-off between delay and collision probability can be achieved with a flexible resource block selection. Monte Carlo simulation results show that with the flexible resource selection scheme average delay can be shortened significantly with only a small compromise in collision probability.
	\end{abstract}
\section{Introduction}

Safety related applications of future connected vehicles are based on the periodic exchange of vehicular status, such as Basic Safety Messages (BSMs) in the DSRC standard. Such BSMs are broadcast periodically and require high reliability and low latency communication. The automotive industry standardized a WiFi extension called IEEE 802.11p (also known as DSRC) to address the communication requirements of safety related applications\cite{vinel20123gpp}.  A recent Notice of Proposed Rule Making (NPRM) issued by the National Highway Transportation Safety Authority (NHTSA) has proposed IEEE 802.11p (DSRC) for the technology to deliver Safety Applications between vehicles in the ITS (5.9 GHz) band – specifically channel 172. The NPRM mentions possibility of using alternative technologies in this band.  

A leading alternative technology is Cellular V2X (C-V2X), which is based on  leveraging and enhancing existing cellular-based LTE features and network elements to facilitate the exchange of V2X messages between vehicles. C-V2X is initially defined as LTE V2X. In 3GPP Release 14, it is designed to operate in several modes: V2V, V2I, V2N and V2P\cite{5g2016case}. The V2V mode overcomes the limitation of the traditional LTE  architecture by introducing a new Sidelink alongside the Downlink and Uplink, which allows vehicles to transmit directly to each other. In 3GPP Release 14, two new modes (modes 3 and 4) specifically designed for V2V communications are introduced. Modes 3 and 4  differ on how they allocate the radio resources. Resources are allocated by the cellular network under Mode 3, while Mode 4 does not require cellular coverage. Vehicles autonomously select their radio resources using a distributed scheduling scheme in Mode 4\cite{molina2017lte}.

In both DSRC and C-V2X systems, the MAC protocol plays a vital role to guarantee a timely communication of BSMs. In DSRC, IEEE 802.11 distributed coordination function MAC protocol has been adopted by the IEEE 802.11p standard for DSRC applications\cite{ieee1999part11}. The MAC performance of DSRC has been well studied.  The authors in \cite{rao2008performance} developed an analytical model to determine the probability of packet collision in the broadcast scenario. The model in \cite{chen2007quantitative} attempted to capture the characteristics of the DSRC safety communications, where broadcasting takes place in an unsaturated network with hidden terminals. In \cite{hassan2011performance}, the authors provided an accurate
analytical model for the MAC protocol of DSRC in unsaturated broadcast networks both with and without hidden terminals. However for the MAC protocol of C-V2X, there is little literature providing analytical models to evaluate the performance.  In \cite{gallo2017unsupervised}, the authors used a case study to evaluate the performance of C-V2X safety communication.  Authors in \cite{gallo2014analytic} proposed an analytical model for C-V2X transmission mode 2 but does not address the case for transmission mode 4, where vehicles choose their resource block in a distributed way. 

In this paper, we mainly focus on the performance analysis of a distributed and semi-persistent resource scheduling scheme which is a simplified version of  transmission mode 4 in C-V2X system. Under the assumption of perfect PHY performance, we develop an analytical model to characterize the scheduling scheme. We then use a fixed point method to approximate the  collision probability and average delay  in a fully connected case. We also consider the hidden terminal issue in partially connected cases. In addition a Monte Carlo simulator is used to verify the results obtained in the analytical model and to conduct more exploration of the  system.  Furthermore, we investigate a flexible resource block selection scheme, with which average delay can be lowered significantly with a slight compromise on collision probability.

This paper is organized as follows. Section \ref{sec:analytical} provides the analytical model we use to characterize the performance of the semi-persistent resource scheduling scheme. Section \ref{sec:monte_result} shows the results of the Monte Carlo simulation and compares them with the results obtained through the analytical model.  Section \ref{sec:conclusion} draws the conclusions.

\section{Analytical modeling of semi-persistent resource scheduling scheme}
\label{sec:analytical}
In this section, we provide the analytical model for the semi-persistent resource scheduling scheme similar to C-V2X transmission Mode 4.  We assume perfect PHY-layer performance to simplify the analysis. More precisely, we assume that any packet sent within a given radius can be heard perfectly if not interfered by others. We consider a system in which vehicles regularly generate BSMs every $T_{tr}$ milliseconds and each $T_{tr}$ millisecond interval contains $N_r$ virtual resource blocks. Each virtual resource block can support the transmission of one BSM. The C-V2X system can be viewed as an example of such framework. In C-V2X system, each Transmission Time Interval (TTI) is $1$ms. In each TTI, there are 50 resource blocks, each comprised of a group of OFDM tones. . Each vehicle broadcasts the BSMs periodically with period $100$ms. The periodic BSMs include $4$ with length 190 Byte and $1$ with length 300 Byte\cite{molina2017lte}. The 190-Byte message requires 17 resource blocks to transmit while the 300-Byte message requires 25 resource blocks to transmit. Note that in each TTI, the resources can support at least 2 BSM transmissions, if we divide the frequency resources in each TTI into two equal parts for vehicles to choose from. In this case,  we can view the resources in 1 TTI as 2 {virtual resource blocks}. Thus, in the C-V2X system, we can assume $T_{tr} = 100$ and $N_r = 200$.

\begin{figure}[h]
	\centering
	\includegraphics[scale=0.3]{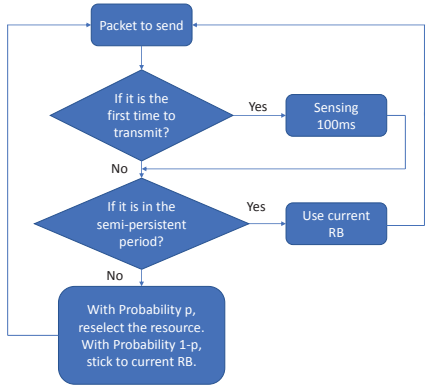}
	\caption{The resource scheduling procedure.}
	\label{fig:flow}
\end{figure}

The semi-persistent resource scheduling scheme is briefly described in Fig. \ref{fig:flow}.  During the semi-persistent period, each vehicle sticks to its current resource block selection, even if a collision occurs, because there is no acknowledgment when broadcasting BSMs. When the current semi-persistent period ends, a vehicle reselects  resource blocks with a certain probability. Each time when a vehicle wants to reselect resources, it will randomly choose from the resource blocks with lowest $20\%$ of transmission power, sensed over the last transmission period. Under the assumption of perfect physical layer performance, the vehicle only senses if the resource block is occupied.  

\subsection{Collision probability}
We use a fixed point model to characterize the collision probability in the fully connected scenario.  Assume the total number of virtual resource blocks in one transmission period is $N_r$, which can be controlled by the length of the transmission period. The total number of vehicles is denoted by $N_v$. We assume that $N_v<N_r$. We use $T_s$ to denote the number of transmission periods in one semi-persistent period. We assume that all vehicles have the same semi-persistent period and are synchronized in terms of the semi-persistent period. When each semi-persistent period ends, the vehicle reselects the virtual resource block with probability $p$.

In the fully connected case, if the BSM from one vehicle collides in the first transmission period in the semi-persistent period, it will collide in the following transmission periods in this semi-persistent period. If the first transmission is successful, so will the subsequent transmissions during the semi-persistent period.. As a result, for any vehicle, we only need to consider the collision probability in the first transmission period in the semi-persistent period. 

In our model, collisions only happens in two cases. One is that the tagged vehicle chooses to reselect its virtual resource block and another vehicle also decides to reselect its virtual resource block at the same time and both vehicles choose the same virtual resource block.  Another case is that the tagged vehicle collides in the previous semi-persistent period and decides not to change virtual resource block selection and at the same time, the vehicle that collides with the tagged one also chooses to stay in current virtual resource block. 

Consider any vehicle in the system as a tagged vehicle. Consider the first case of collision. In one transmission period, the probability that $n$ $(n\ge1)$ out of $N_v-1$ vehicles decide to reselect their virtual resource  is 
\begin{equation}
\label{re_sel}
P_r(n) = C(N_v-1,n)\left(\frac{p}{T_s}\right)^n\left(1-\frac{p}{T_s}\right)^{N_v-1-n}.
\end{equation}

Next we calculate the expected number of idle resource blocks that vehicles can sense during the transmission period. Based on our assumption that $N_v<N_r$, the probability that collision happens among the messages from more than 2 vehicles is low. Thus the expected  number of idle resource blocks seen by the vehicles can be approximated by
\begin{equation}
\label{Nidle}
N_{idle} = N_r-N_v+P_c(N_v-1)/2,
\end{equation}
where $P_c$ is the overall collision probability we need to calculate. When $n$ vehicles decide to change virtual resource selection at the same time of the tagged vehicle, the probability that the choice of at least one of the $n$ vehicles collides with the tagged vehicle is
\begin{equation}
\label{col}
P_s(n) = 1-\left(\frac{N_{idle}-1}{N_{idle}}\right)^n.
\end{equation}

As a result the collision probability in case 1 is
\begin{equation}
\label{col_prob}
P_c^I=  \sum\limits_{n=1}^{N_v-1} P_r(n)P_s(n).
\end{equation}  

We can substitute (\ref{re_sel}) and (\ref{col}) into (\ref{col_prob}) and use binomial theorem to get the following equation:
\begin{equation}
\label{fix_point}
P_c^I = \left[1-\left(1-\frac{p}{T_s N_{idle}}\right)^{N_v-1}\right].
\end{equation}

For case 2,  it is difficult to identify the exact number of transmissions that collide with the tagged one. But as we stated before for the expected number of idle resource blocks, most collision happens between two vehicles.	As a result, the collision probability in case 2 can be approximated as 
\begin{equation}
\label{P_c2}
P_c^{II} = (1-p)P_c,
\end{equation}
where $P_c$ is the overall collision probability. Consequently, the collision probability $P_c$ can be written as 
\begin{equation}
\label{P_c}
P_c = p P_c^I+ (1-p)P_c^{II}.
\end{equation}

By substituting $P_c^I$ and $P_c^{II}$ into (\ref{P_c}), we can find the following fixed point problem
\begin{equation}
\label{eqn:fix}
P_c = \frac{1}{2-p}\left[1-\left(1-\frac{p}{T_s N_{idle}}\right)^{N_v-1}\right],
\end{equation}
where $N_{idle}$ is defined in (\ref{Nidle}) and depends on $P_c$. .
	
\subsection{Average delay}
Delay is defined as the time gap between currently received BSM and the BSM that the vehicle should have received since vehicle's last received BSM.
	For semi-persistent scheduling, delay can be caused by the initial delay and collision, i.e. 
	\begin{equation}
	\label{eqn:delay}
	D_{\rm cv2x} = D_{\rm ini} + D_{\rm col}.
	\end{equation}
	
	The initial delay is caused by the random selection of the resource block. After the channel monitoring period (under our assumption equal to the transmission period of 100ms) the vehicle selects a virtual resource block uniformly randomly in the next transmission period. We will observe later that the choice of the selection window has interesting implications on the latency-collision probability trade-off. This initial delay can be viewed as a uniform random variable in $\left[0,T_{tr}\right]$, where $T_{tr}$ is the transmission period in ms.  Thus we have
	\begin{equation}
	E[ D_{\rm ini}] = \frac{T_{tr}}{2}. 
	\end{equation}
	
	The collision delay is caused by the packet loss when collision happens. In the fully connected case, once collision happens, it lasts for the whole semi-persistent period, thus $T_s$ BSMs are lost and that will cause ${T_s}{T_{tr}}$ in delay. As a result, we can consider the collision in one semi-persistent period as one single collision with delay ${T_s}{T_{tr}}$. We can view this period as a combined one with 1 transmission and 1 collision instead of $T_s$ transmissions and $T_s$ collisions. Since the overall collision probability is $P_c$, the probability that the combined collision happens can be expressed as
	\begin{equation}
	\label{eqn:equivalent_col_prob}
	P_c^{com} = \frac{P_c}{T_s-(T_s-1)P_c}.
	\end{equation}
	
	To simplify the model, we assume that combined collision slot happens independently  from slot to slot.. The collision delay then follows a geometric distribution and as a result the expected delay is given by 
	\begin{small}
			\begin{equation}
		E[D_{\rm col}] = \sum\limits_{i=1}^{\infty} {iT_s}{T_{tr}}\left(P_c^{com}\right)^i\left(1-P_c^{com}\right) = \frac{T_sT_{tr}P_c^{com}}{\left(1-P_c^{com}\right)},
		\end{equation}
	\end{small}

	where $P_c^{com}$ is defined in (\ref{eqn:equivalent_col_prob}).

\subsection{Hidden terminal issue}
When not all the vehicles can hear each other, i.e., in a partially connected topology, we need to consider the hidden terminal issue. We assume that the vehicles are uniformly distributed and  the vehicle density on the high way is denoted by $\beta$ (vehicles/km) and the transmission range of each vehicle is $R$. In such a case, if we choose one arbitrary vehicle  as the coordinate origin, then the transmission range is $[-R,R]$ and possible hidden terminals appear in range $[-2R,-R]$ and $[R,2R]$.

% as illustrated in Fig. \ref{fig:ill}. 
%\begin{figure}[h]
%	\centering
%	\includegraphics[scale=0.3]{pic/ill}
%	\caption{Partially connected model illustration}
%	\label{fig:ill}
%\end{figure}

Thus the expected number of vehicles in the transmission range of the given vehicle is $N_{tr} = 2\beta R$ and the number of possible hidden terminals is $N_{ht} = 2\beta R$. To calculate the collision probability with hidden terminal issues, we first consider the case that all possible hidden terminals are silent. This case has been described in previous section. The collision probability can be calculated by solving the  fixed point problem in (\ref{eqn:fix}). The packet delivery ratio without hidden terminal is thus 
\begin{equation}
\label{delivery}
P_{del} = 1-P_c,
\end{equation} 
where $P_c$ is the collision probability in the fully connected case. However the packet may collide with hidden terminals if any of the vehicles in the hidden terminal range use the same virtual resource block. Since the vehicles in the hidden terminal range also uses the same resource selection scheme, they can see the resource blocks selected by users in their own transmission range. On average, vehicles in the hidden terminal range of the tagged vehicle can see at most $\beta R$ virtual resource blocks used by the vehicles in the transmission range of the tagged vehicle. For simplification, we assume the vehicles in the hidden terminal range do not use the $\beta R$ virtual resource blocks and randomly chooses from the rest $N_r-\beta R$ virtual resource blocks. The probability that a single vehicle in the hidden terminal range does not collide with the tagged vehicle is then
\begin{equation}
\label{eqn:psingle}
P_{\rm single} = \frac{N_r-\beta R-1}{N_r-\beta R}.
\end{equation}
To guarantee the delivery of the message, all vehicles in the hidden terminal range should not use the same virtual resource block. As a result the collision probability considering the hidden terminal issue becomes
\begin{equation}
\label{collision_ht}
P_{c}^{ht} = 1-(1-P_c)\left(P_{\rm single}\right)^{N_{ht}}.
\end{equation} 

Note that the collision probability is that seen  from the transmitter side. As long as one collision happens while other vehicles received the message, it is still considered as a collision. Another way to think of collisions is from the receiver side, which we refer to as the packet error ratio. Packet error ratio is defined as the following. 
\[PER = \frac{{\rm number\; of\;  collided \; messages}}{\rm total \; number\;  of\;  messages}.\]
This characterizes the ratio of missed messages in all the messages that the vehicle should have received. In the fully connected case, packet error ratio is equal to collision probability. But in the partially connected case, the packet error ratio is different from the collision probability. It can be expressed as
\begin{small}
\begin{eqnarray}
PER & =&  1- 2\int_{0}^{R}\frac{1}{2R}\left(1-P_c\right)P_{\rm single}^{\beta r} dr \nonumber\\
&=& 1-\frac{1}{\beta R \ln \left(P_{\rm single}\right)}\left(1-P_c\right)\left[P_{\rm single}^{\beta R} -1\right],\nonumber
\end{eqnarray}
\end{small}
where $P_{\rm single}$ is defined in (\ref{eqn:psingle}) and $P_c$ is the collision probability calculated in the fully connected case.	

The average delay can be calculated in the same way as the fully connected case. The only difference is that in the partially connected case, we should use packet error ratio for the calculation.

\section{Simulation results}
\label{sec:monte_result}
We conduct Monte Carlo simulation under static vehicle distribution to study the accuracy  of the analytical results we derived in the previous section. In each experiment, we assume that the number of vehicles in the system is fixed and the location of vehicles remains the same. The location of each vehicle is determined by a uniformly distributed random variable with the maximum value to be the length of road. The transmission time period we use in the experiment is $ T_{tr}=100$ms, i.e., every vehicle generates a BSM every 100ms. Each experiment runs for 2000 seconds and the same experiment is conducted 10 times to average the results.

\subsection{Fully connected case}	
\begin{figure}[htbp]
	\centering
	\subfigure[ Collision probability ]{
		\label{fig:delay_cv2x} %% label for first subfigure
		\includegraphics[width=1.6in]{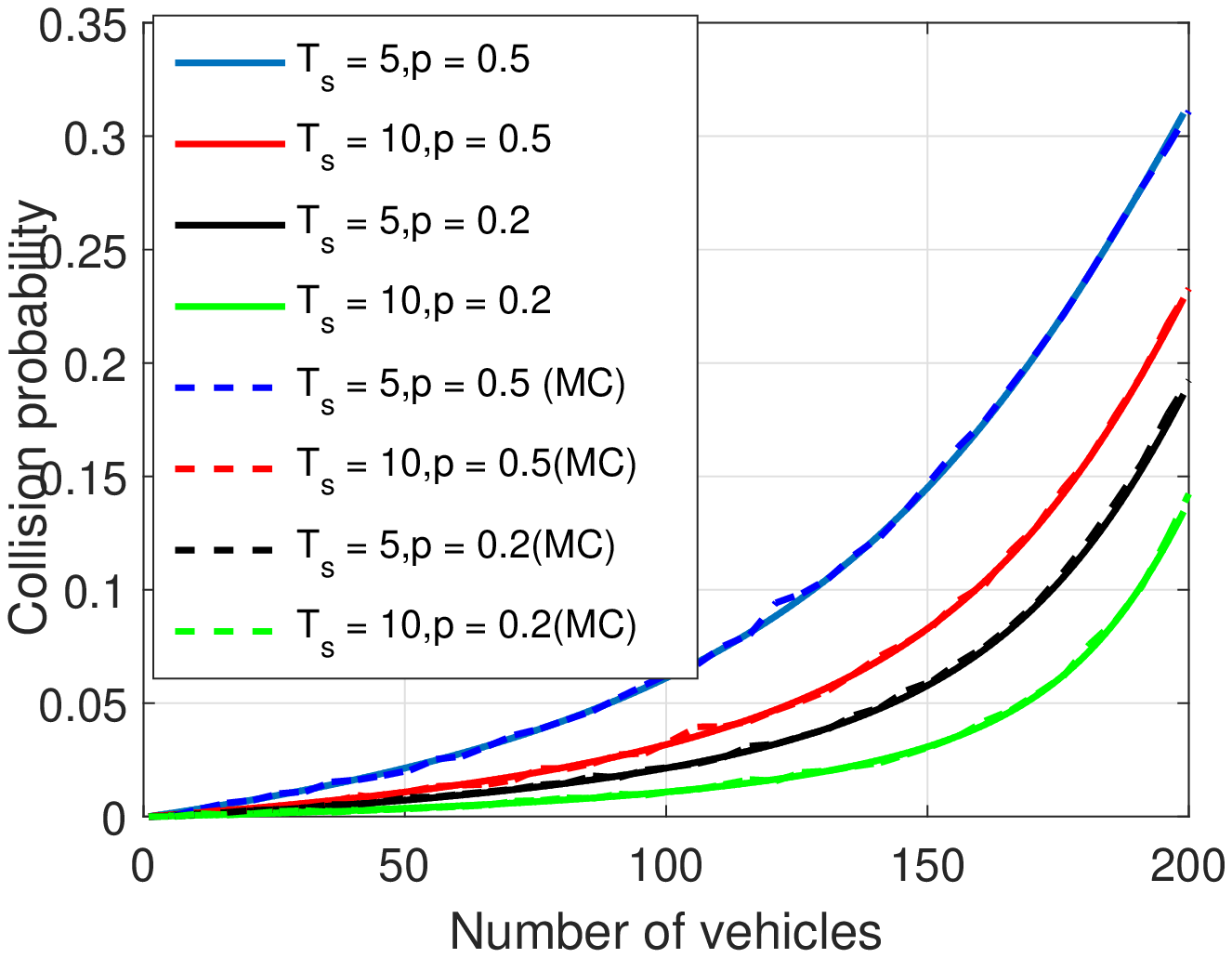}}
	%\hspace{0.1in}
	\subfigure[Average delay]{
		\label{fig:delay_dsrc} %% label for second subfigure
		\includegraphics[width=1.6in]{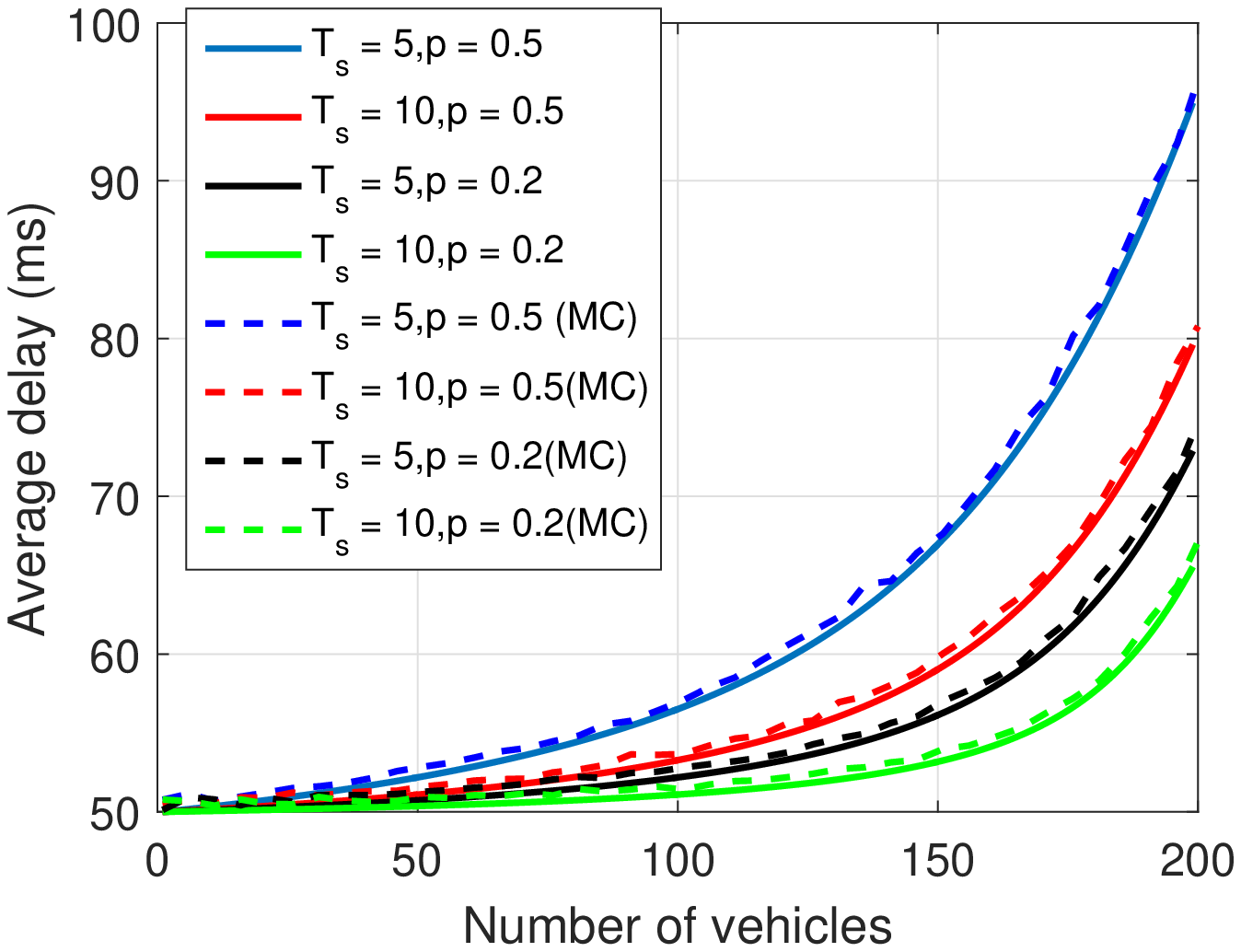}}
	\caption{Simulation results for fully connected network.}
	\label{fig:monte} %% label for entire figure
\end{figure}

We first see if the collision probability and average delay calculated in the analytical model is accurate in the fully connected case. Simulation results on collision probability and average delay with different semi-persistent period $T_s$ and resource reselection probability $p$ are shown in Fig. \ref{fig:monte}. In both figures, solid curves denote the values calculated from analytical model and dashed curves show the results from Monte Carlo simulation. We can see that both collision probability and delay simulation matches quite well with the analytical results under different parameters. In this model, the value of $\frac{p}{T_s}$ indicates the probability that a vehicle is going to change resource block selection. We can see from the Monte Carlo results that when $\frac{p}{T_s}$ decreases, collision probability decreases. This implies that one may  restrict the frequency of resource reselection to lower the collision probability in uncongested networks ($N_v<N_r$). 

\begin{figure}[htbp]
	\centering
	\includegraphics[scale=0.30]{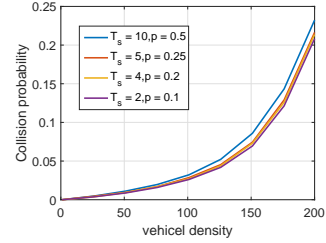}
	\caption{Collision probability when $p/T_s$ is fixed.}
	\label{fig:fully_connected}
\end{figure}

 It is also interesting to see when we fix $\frac{p}{T_s}$ and change $p$, how will that affect the collision probability. Monte Carlo simulation results are shown in Fig. \ref{fig:fully_connected}. We can see that if $p/T_s$ is fixed when $p$ increases, collision probability increases. From an analytical point of view, the collision probability comes from the  fixed point of (\ref{eqn:fix}). Since the LHS is increasing function  in $P_c$ and the RHS is decreasing in $P_c$, when $p/T_s$ is fixed and $p$ increases, the fixed point should also increase. This result implies that one may want to  choose a smaller $p$ when  fixing $\frac{p}{T_s}$.

\subsection{Partially connected case}
\begin{figure}[htbp]
	\centering
	\subfigure[ Collision probability \& Packet error ratio ]{
		\label{fig:mc_ht_col} %% label for first subfigure
		\includegraphics[width=1.6in]{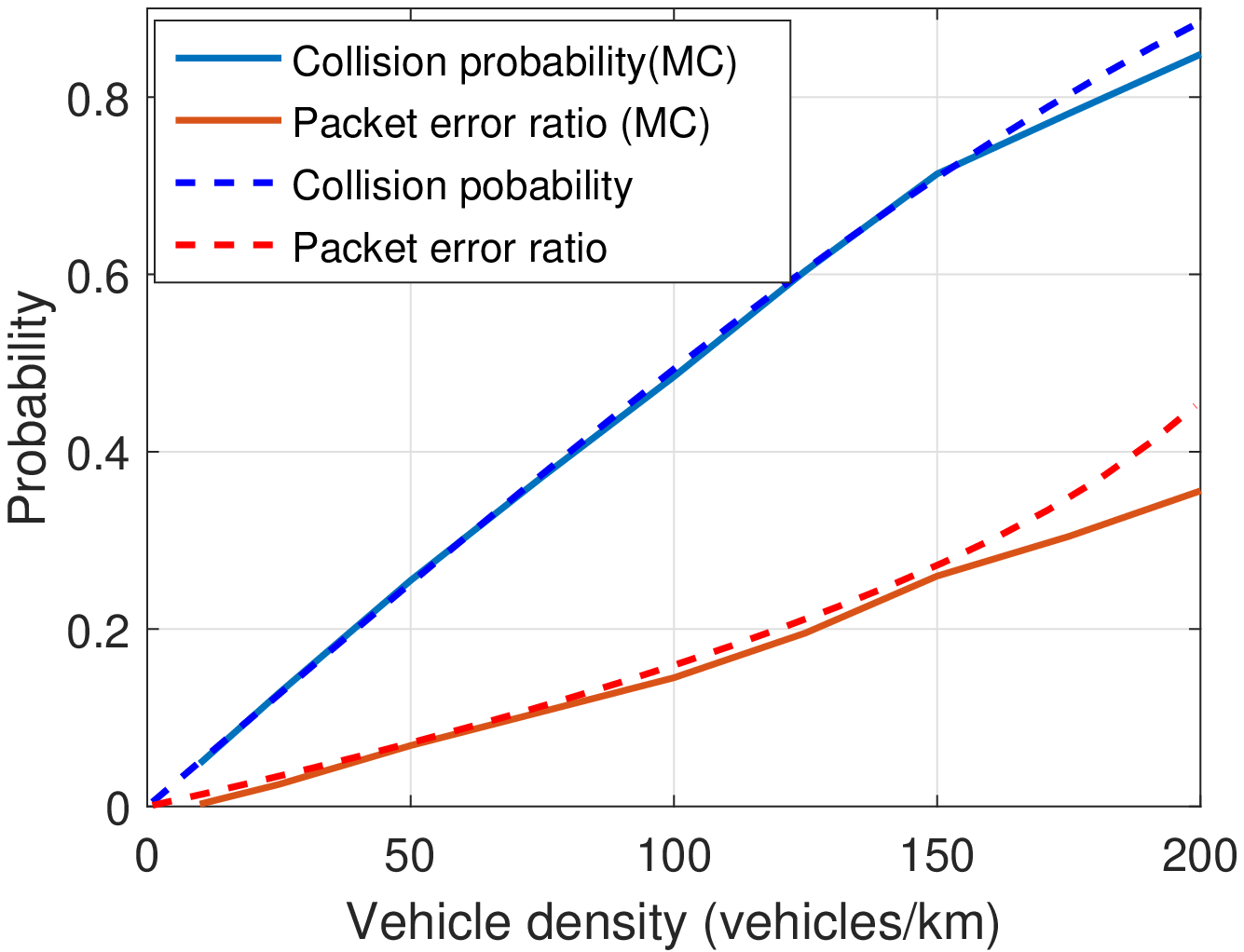}}
	%\hspace{0.1in}
	\subfigure[Average delay]{
		\label{fig:mc_ht_delay} %% label for second subfigure
		\includegraphics[width=1.6in]{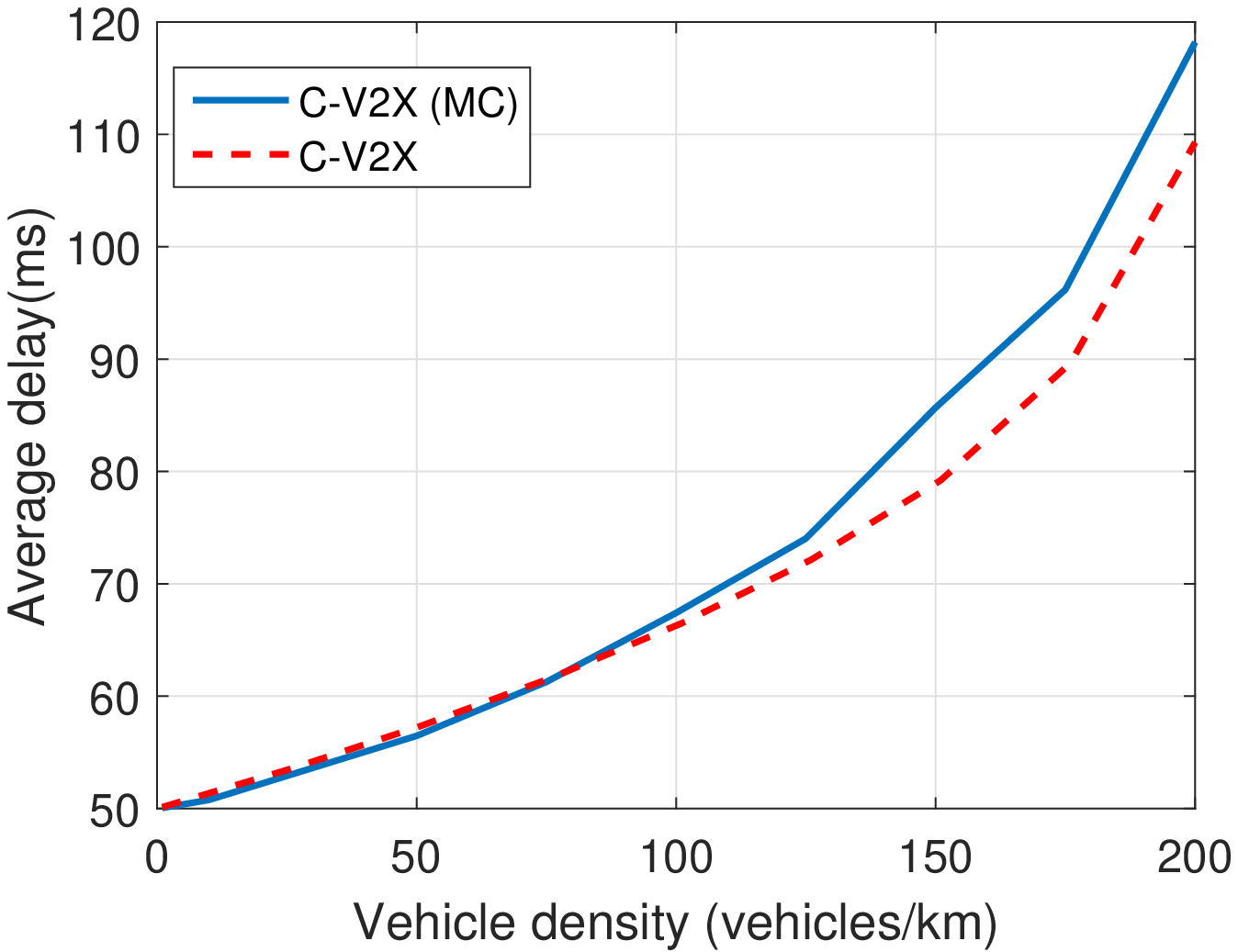}}
	\caption{Simulation results for partially connected network.}
	\label{fig:mc_ht} %% label for entire figure
\end{figure}

We also verify the analytical results in the partially connected case.  In the simulation, we set the parameter $T_s =10$, $p=0.2$ and transmission range $R=500$m. The results of collision probability and packet error ratio and average delay are shown in Fig. \ref{fig:mc_ht}. It can be seen that the analytical results and simulation results match quite well with each other especially when traffic load is low. When traffic load is high, there is some difference but the results are still close. A main reason for the difference is that our calculation decouples the impact of hidden terminals and the resource scheduling scheme. When vehicle density is relatively high, these two factors may impact each other, which leads to the difference.

\begin{figure}[h]
	\centering
	\includegraphics[scale=0.3]{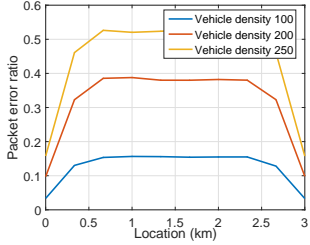}
	\caption{Performance with different location in 3km road. }
	\label{fig:diff_location} %% label for entire figure
\end{figure}

In the partially connected case, packet error ratio may change with location if the road length is not infinity. We consider a case where the total length of road is 3km, where the transmission range is 500m. We show how packet error ratio changes with location under vehicle density 100, 200 and 250 in Fig. \ref{fig:diff_location}. We can see that in the center, packet error ratio is higher than that in the edge, because vehicles on the edge suffers less from the hidden terminal issue.  

\subsection{Delay-Optimized resource block selection}
As is seen in previous figures, even though the traffic is light in the system, average delay is still around $50$ms, due to our assumption that a virtual resource block is chosen uniformly within the whole subsequent transmission period. Since vehicles can only transmit the packet after the packet arrives, the initial delay $D_{ini}$ in (\ref{eqn:delay}) is a major part in the average delay. However it is possible to reduce this by allowing vehicles to choose the closest idle virtual resource block in the current transmission period.  With such a selection scheme, we may expect slightly more collisions, because it is more likely for vehicles to select the same virtual resource block. However average delay can be shortened.  
\begin{figure}[htbp]
	\centering
	\subfigure[ Collision probability ]{
		\label{fig:oprimized_prob} %% label for first subfigure
		\includegraphics[width=1.6in]{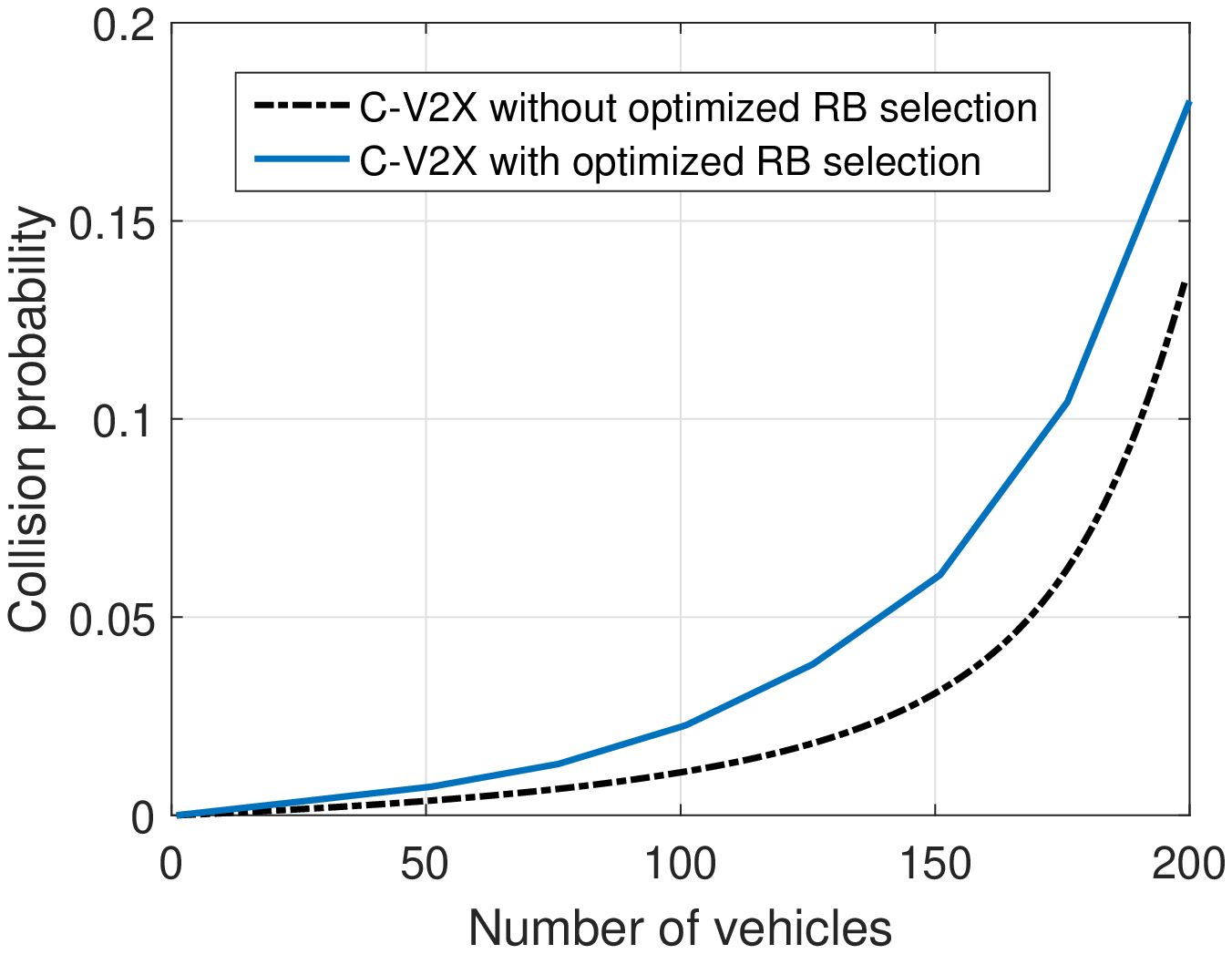}}
	%\hspace{0.1in}
	\subfigure[Average delay]{
		\label{fig:optimized_delay} %% label for second subfigure
		\includegraphics[width=1.6in]{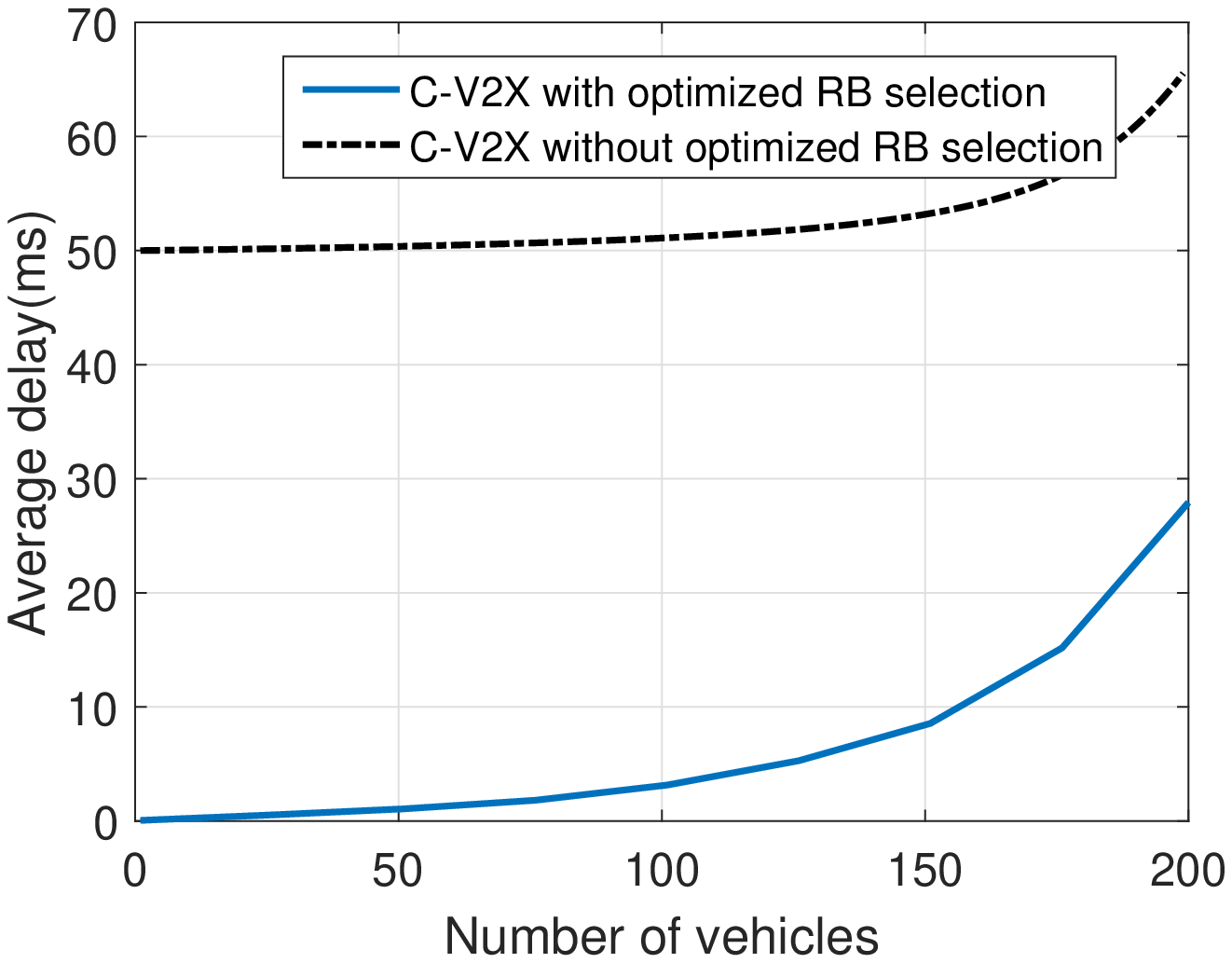}}
	\caption{ Comparison between different resource selection approaches in fully connected case.}
	\label{fig:optimized} %% label for entire figure
\end{figure}

\begin{figure}[htbp]
	\centering
	\subfigure[ Packet error ratio]{
		\label{fig:cv2x_ht_per_optimized} %% label for first subfigure
		\includegraphics[width=1.6in]{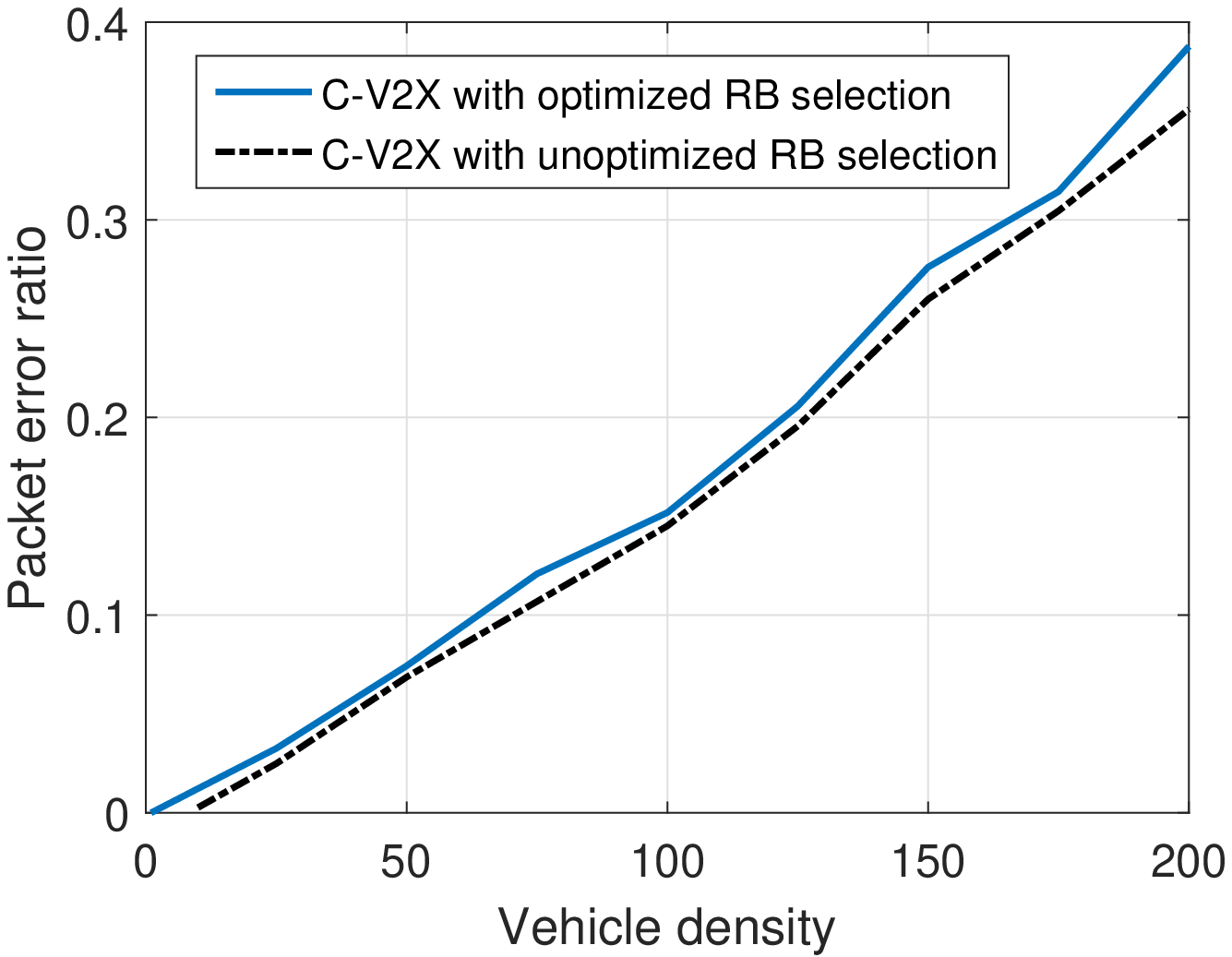}}
	%\hspace{0.1in}
	\subfigure[Average delay]{
		\label{fig:cv2x_ht_delay_optimized} %% label for second subfigure
		\includegraphics[width=1.6in]{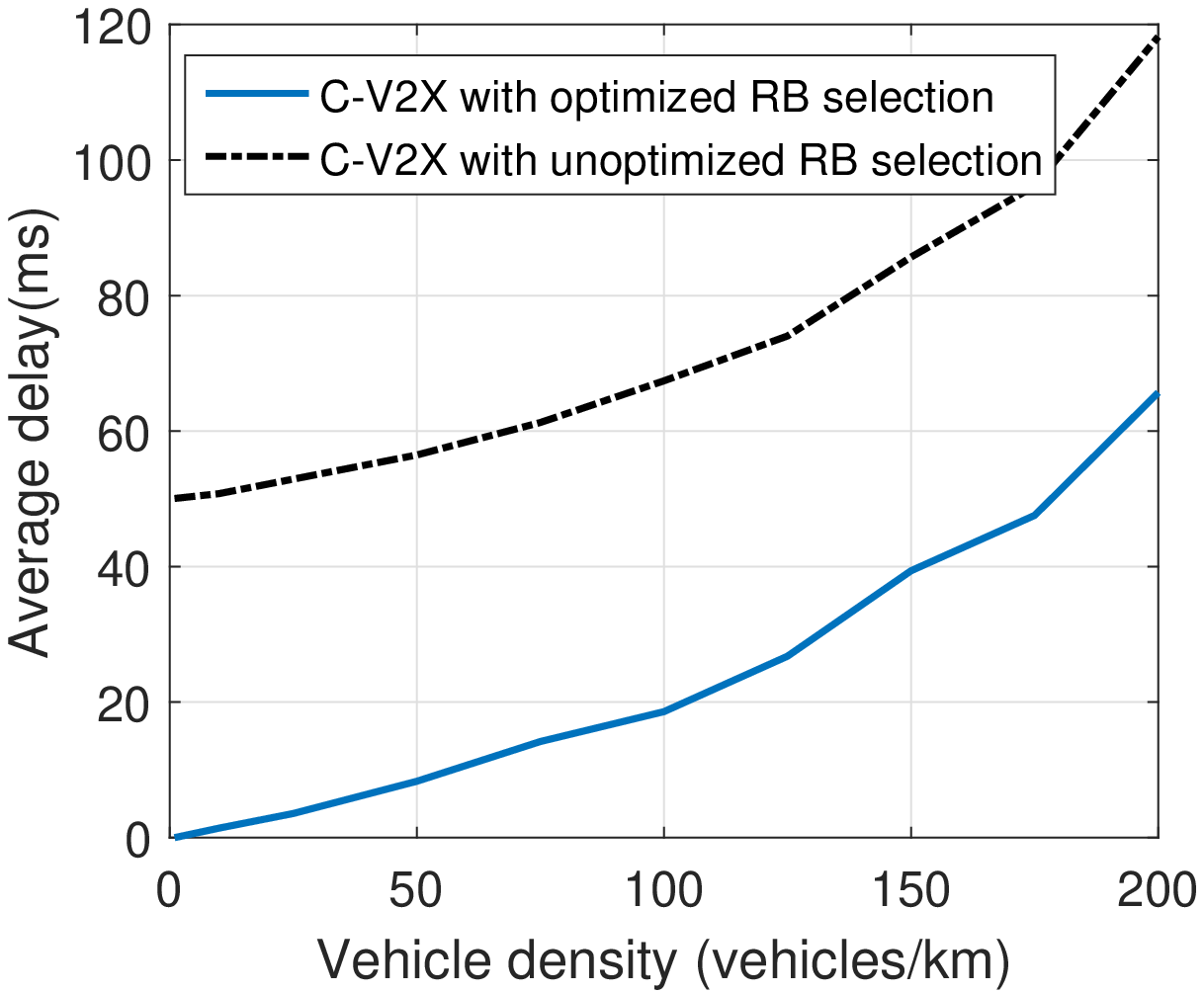}}
	\caption{Comparison between different resource selection approaches in partially connected case.}
	\label{fig:ht_optimized} %% label for entire figure
\end{figure}

We conduct Monte Carlo simulation for this delay-optimized resource scheduling scheme in both fully and partially connected cases. Results are shown in Fig. \ref{fig:optimized} and Fig. \ref{fig:ht_optimized}.  We can see that in both cases, with very small sacrifice on packet error ratio, delay is shortened significantly.

\section{Conclusion}
\label{sec:conclusion}
In this paper, we focused on the performance analysis of a semi-persitent resource scheduling scheme which represents a simplified version of C-V2X system transmission Mode 4. We built an analytical model and used a fixed point method to analyze the collision probability and average delay in both fully connected and partially connected cases. In addition we used Monte Carlo simulation to verify the results obtained in the analytical model. Results showed that our analytical model can give a good estimation of the collision probability and average delay. We also investigated a delay-optimized resource selection scheme with which average delay can be shortened significantly. 

In our work, we assume a perfect PHY performance, introducing RF modeling is a possible extension. Also we only investigated the performance of C-V2X system under a static setting, i.e., all vehicles are assumed to be static. It would be interesting to look at the transients when vehicles moves between areas with different traffic. Furthermore, it is possible to compare the performance with DSRC system to find the advantages and disadvantages of these two technologies.

\bibliographystyle{plain}
\bibliography{mybib}

\begin{thebibliography}{1}

\bibitem{5g2016case}
5G~Automotive Association et~al.
\newblock The case for cellular {V2X} for safety and cooperative driving.
\newblock {\em 5GAA Whitepaper, Nov}, 2016.

\bibitem{chen2007quantitative}
Xianbo Chen, Hazem~H Refai, and Xiaomin Ma.
\newblock A quantitative approach to evaluate {DSRC} highway inter-vehicle
  safety communication.
\newblock In {\em Global Telecommunications Conference, 2007. GLOBECOM'07.
  IEEE}, pages 151--155. IEEE, 2007.

\bibitem{gallo2017unsupervised}
Laurent Gallo and Jerome Haerri.
\newblock Unsupervised {Long-Term} {Evolution} {Device-to-Device}: A case study
  for safety-critical {V2X} communications.
\newblock {\em IEEE Vehicular Technology Magazine}, 2017.

\bibitem{gallo2014analytic}
Laurent Gallo and J{\'e}rome Harri.
\newblock {\em Analytic performance comparison of unsupervised {LTE} {D2D} and
  {DSRC} in a {V2X} safety context}.
\newblock PhD thesis, EURECOM, 2014.

\bibitem{ieee1999part11}
IEEE 802.11~Working Group et~al.
\newblock Part11:{ Wireless} {LAN} medium access control {(MAC)} and physical
  layer {(PHY)} specifications.
\newblock {\em ANSI/IEEE Std. 802.11}, 1999.

\bibitem{hassan2011performance}
Md~Imrul Hassan, Hai~L Vu, and Taka Sakurai.
\newblock Performance analysis of the {IEEE} 802.11 {MAC} protocol for {DSRC}
  safety applications.
\newblock {\em IEEE Transactions on Vehicular Technology}, 60(8):3882--3896,
  2011.

\bibitem{molina2017lte}
Rafael Molina-Masegosa and Javier Gozalvez.
\newblock {LTE-V} for {Sidelink} {5G} {V2X} vehicular communications: A new
  {5G} technology for short-range {Vehicle-to-Everything} communications.
\newblock {\em IEEE Vehicular Technology Magazine}, 12(4):30--39, 2017.

\bibitem{rao2008performance}
Ashwin Rao, Arzad Kherani, and Anirban Mahanti.
\newblock Performance evaluation of 802.11 broadcasts for a single cell network
  with unsaturated nodes.
\newblock In {\em International Conference on Research in Networking}, pages
  836--847. Springer, 2008.

\bibitem{vinel20123gpp}
Alexey Vinel.
\newblock {3GPP} {LTE} versus {IEEE} {802.11p/WAVE}: {Which} technology is able
  to support cooperative vehicular safety applications?
\newblock {\em IEEE Wireless Communications Letters}, 1(2):125--128, 2012.

\end{thebibliography}

\end{document}